\documentclass[a4paper, 12pt]{article}

\usepackage[a4paper, margin=1in]{geometry}
\usepackage{mathtools}
\usepackage{graphicx}
\usepackage{booktabs}
\usepackage{amssymb}
\usepackage{algpseudocode}
\usepackage{breqn}
\usepackage[svgnames]{xcolor}
\usepackage[pdfusetitle,colorlinks,plainpages=false,citecolor=violet,linkcolor=violet]{hyperref}
\usepackage[round]{natbib}
\usepackage[ruled]{algorithm}
\usepackage{textcomp}
\graphicspath{{./Figures/}}

\newcommand{\argmin}{\operatornamewithlimits{argmin}}
\newcommand\ddfrac[2]{\frac{\displaystyle #1}{\displaystyle #2}}

\begin{document}
\title{Granular clustering of \textit{de novo} protein models}

\author{Dmytro Guzenko and Sergei V. Strelkov\,$^{*}$ \\ {\small Department of Pharmaceutical and Pharmacological Sciences} \\ {\small KU Leuven, Leuven 3000, Belgium.}\\ {\small $^\ast$To whom correspondence should be addressed.}}

\date{}
\maketitle

\section*{Abstract}
Modern algorithms for \textit{de novo} prediction of protein structures typically output multiple full-length models (decoys) rather than a single solution. Subsequent clustering of such decoys is used both to gauge the success of the modelling and to decide on the most native-like conformation. At the same time, partial protein models are sufficient for some applications such as crystallographic phasing by molecular replacement (MR) in particular, provided these models represent a certain part of the target structure with reasonable accuracy. Here we propose a novel clustering algorithm that natively operates in the space of partial models through an approach known as granular clustering (GC). The algorithm is based on growing local similarities found in a pool of initial decoys. We demonstrate that the resulting clusters of partial models provide a substantially more accurate structural detail on the target protein than those obtained upon a global alignment of decoys. As the result, the partial models output by our GC algorithm are also much more effective towards the MR procedure, compared to the models produced by existing software. The source code is freely available at https://github.com/biocryst/gc

\section{Introduction}

Protein \textit{de novo} 3-dimensional (3D) structure prediction involves extensive sampling of the conformation space in search of the near-native low energy state. The large number of decoys produced makes it impossible to inspect and interpret the results manually. Structural clustering is a widely used tool for post-processing of \textit{de-novo} folded decoys (\citealp{leaver2011rosetta3}, \citealp{Zhang:Tasser}). It exploits the idea that frequently sampled low-energy conformations are more likely to represent the native structure than the lone lowest-energy decoy (\citealp{Shortle15091998}). 

Clustering algorithms require a dissimilarity measure between any two objects. This function involves a superposition of the structures that optimises certain score, most typically a root-mean-square deviation (RMSD) of atomic positions minimised with the Kabsch algorithm (\citealp{kabsch1976solution}). Clearly, a single superposition of full-length models often fails to reveal a complete information on their local similarities. An obvious model situation is a protein consisting of two domains connected by a flexible linker. Such protein can accept a multitude of conformations that are globally very different, even though the conformations of individual domains remain the same. Assessment of local model quality independently of domain motions has long been implemented during the Critical Assessment of Methods for Structure Prediction of Proteins (CASP) competition, with specialised metrics continuously developed (\citealp{Zemla01072003}, \citealp{Mariani01112013}). However, clustering algorithms routinely used to post-process decoys generated by \textit{de-novo} protein folding are still based on single-alignment approaches.

Lack of sensitivity for local similarities inherently limits the capabilities of the cluster analysis towards extracting useful information from the pool of decoys. To overcome this difficulty, one can generate more decoys, hoping that the correct global fold reveals itself as a statistically significant cluster. This is a viable approach if the aim is to obtain an accurate full-length model, but it requires significant computational resources and specific optimisations to handle large distance matrices (\citealp{zhang2004spicker}), since hundreds of thousands of decoys are not uncommon. 

Here we describe a new method to obtain partial protein models which is based on the granular clustering (GC) paradigm (\citealp{Pedrycz:granular}). It outputs substructures that are similar in a sufficiently large number of decoys, without prior assumptions on the modelling accuracy or the substructure size. Clusters of full-length models are the ultimate possibility, making this method complementary to the clustering based on global alignment, \textit{i.e.} working bottom-up, \textit{vs}. top-down, towards the same goal.

Our method is especially useful in applications where partial models covering different fragments of the protein sequence, possibly providing alternative conformations, are sufficient. An important example of such an application is solving the "phase problem" in X-ray crystallography by molecular replacement (MR). Traditionally, the MR procedure required the availability of an experimentally determined protein structure that is sufficiently homologous to the target protein (\citealp{rossmann1990molecular}). More recently, as the methods for \textit{ab initio} structure prediction continued to improve, the possibility to use predicted (partial) structures in MR searches has been demonstrated (\citealp{das2008macromolecular}, \citealp{zhang2009protein}, \citealp{Bibby:tz5014}). Here we show that our partial models obtained through GC are twice as effective in the MR procedure, compared to models prepared using existing approaches from the same initial pool of decoys. Further applications of GC of protein models may include local contacts prediction, non-linear structural motifs discovery, or generation of custom libraries of structural fragments (\citealp{rohl2004protein}).

\section{System and methods}

\subsection{Principles of granular clustering}
We formulate the problem within the granular computing paradigm (\citealp{Jing:granularreview}), which is particularly suited for our bottom-up approach towards obtaining clusters of partial models. The general principles of GC (\citealp{Pedrycz:granular}) are as follows: 
\begin{itemize}
\item Primitive \textit{information granules} are created from the input data elements; these are subsets of the data that can be directly aggregated by a specific property.
\item Clustering is carried out by \textit{growing} information granules -- iteratively merging granules that have significant overlap;
\item Clustering is stopped when enough data \textit{condensation} is achieved.
\end{itemize}

The criteria for information granulation, granule merging and data condensation are not generally defined and are specific for a particular application. 

\subsection{Granular clustering of protein decoys}
We start by defining the granular protein clustering as a search problem, \textit{i.e.}, in terms of initial state, production rule and the goal state. Organisation of the search itself will be discussed in the next section.

Let us consider a set of $K$ decoys for a given protein containing $N$ amino acid residues. Then $\mathbf{R}=\{1,..., N\}$ is a set of numbered residues of this protein and $\mathbf{M}=\{1,..., K\}$ is a set of numbered decoys. 
Cluster $C$, which may contain only some stretches along the sequence rather than the complete protein, and represented by only some of the available decoys, is defined as a pair $C=(R,M), R \subseteq \mathbf{R}, M \subseteq \mathbf{M}$.

Let $\operatorname{V}(R,M)$ be a scoring function of a cluster, subject to minimisation. An obvious example of such function is $\operatorname{RMSD}(R,M)$ -- the average root-mean-square deviation between all pairs of decoys $m_i, m_j \in M, i \ne j$ evaluated on the optimal superposition of C$\alpha$ atoms of residues $R$.

A cluster is \textit{valid} with respect to parameters $(v,s)$ if it consists of  $s$ or more models, superposition of which by the given residues $R$ produces score of at most $v$.

\begin{equation}
\operatorname{Valid}(R,M|v,s) \iff
\begin{array}{l}
\left\vert{M}\right\vert \geq s \\
\operatorname{V}(R,M)\leq v
\end{array}
\label{eq:valid}
\end{equation}

A cluster is \textit{saturated} with respect to parameters $(v,s)$, if no further models can be added to the cluster without breaking its validity.
\begin{equation}
\operatorname{Saturated}(R,M|v,s) \iff
\begin{array}{l}
\operatorname{Valid}(R,M|v,s) \\
\nexists M' \supset M: \operatorname{Valid}(R,M'|v,s)
\end{array}
\label{eq:saturated}
\end{equation}

Let $\mathbf{C}(R,M|v,s)$ be a set of all saturated clusters with respect to parameters $v,s$, further denoted as $\mathbf{C}$ for brevity.

\textit{Initial state}. The set of all clusters $\mathbf{C}$ contains all one-residue segments with the entire set of models as support, i.e. 
\begin{equation}
(\{i: i \in \mathbf{R}\},\mathbf{M}) \in \mathbf{C}
\label{eq:initial_single}
\end{equation}
Since each of these clusters contain all models to begin with, no further models can be added. Thus the saturation condition is naturally satisfied.

\textit{Production rule}. Two distinct clusters can be combined (\textit{i.e.} condensed) if they share enough supporting models and stay below a given score limit. In order to ensure eventual algorithm termination, we require that each input cluster for the production rule contains at least one residue not found in another cluster.
Let $C_1 = (R_1,M_1) \in \mathbf{C}$, $C_2 = (R_2,M_2) \in \mathbf{C}$, where $R_1$ and $R_2$ are not subsets of each other, then a set of clusters produced by $C_1$ and $C_2$ is defined as:

\begin{equation}
\label{eq:prod}     
\operatorname{Prod}(C_1,C_2|v,s) = \big\{(R,M) \in \mathbf{C} : R=R_1 \cup R_2, M \subseteq M_1 \cap M_2 \big\}
\end{equation}

Note that there may be several clusters produced by one pair of inputs. This may happen when two substructures jointly adopt alternative conformations that are sufficiently supported by the pool of decoys. For example, parallel and antiparallel configurations of $\alpha$-helical chains are drastically different structural arrangements, yet they may be regulated by subtle changes in hydrophobic core packing energy (\citealp{malashkevich2015switch}).

A cluster is \textit{terminal} if no new clusters can be produced using it as one of the inputs. $Prod(C_T,C|v,s)= \varnothing, \forall C \in \mathbf{C}$. The \textit{goal state} of the granular clustering is then defined by finding all terminal clusters.

The clustering problem is now formulated as a classical combinatorial search, which enables application of the wealth of methods developed for this purpose. As a proof of concept, we implemented a search by greedy heuristic with a number of simplifications, which allows reaching suboptimal yet evidently useful results within a short computational time.

\section{Algorithm}
\begin{figure}
\centering
\label{fig:01}
\includegraphics[width=0.5\textwidth]{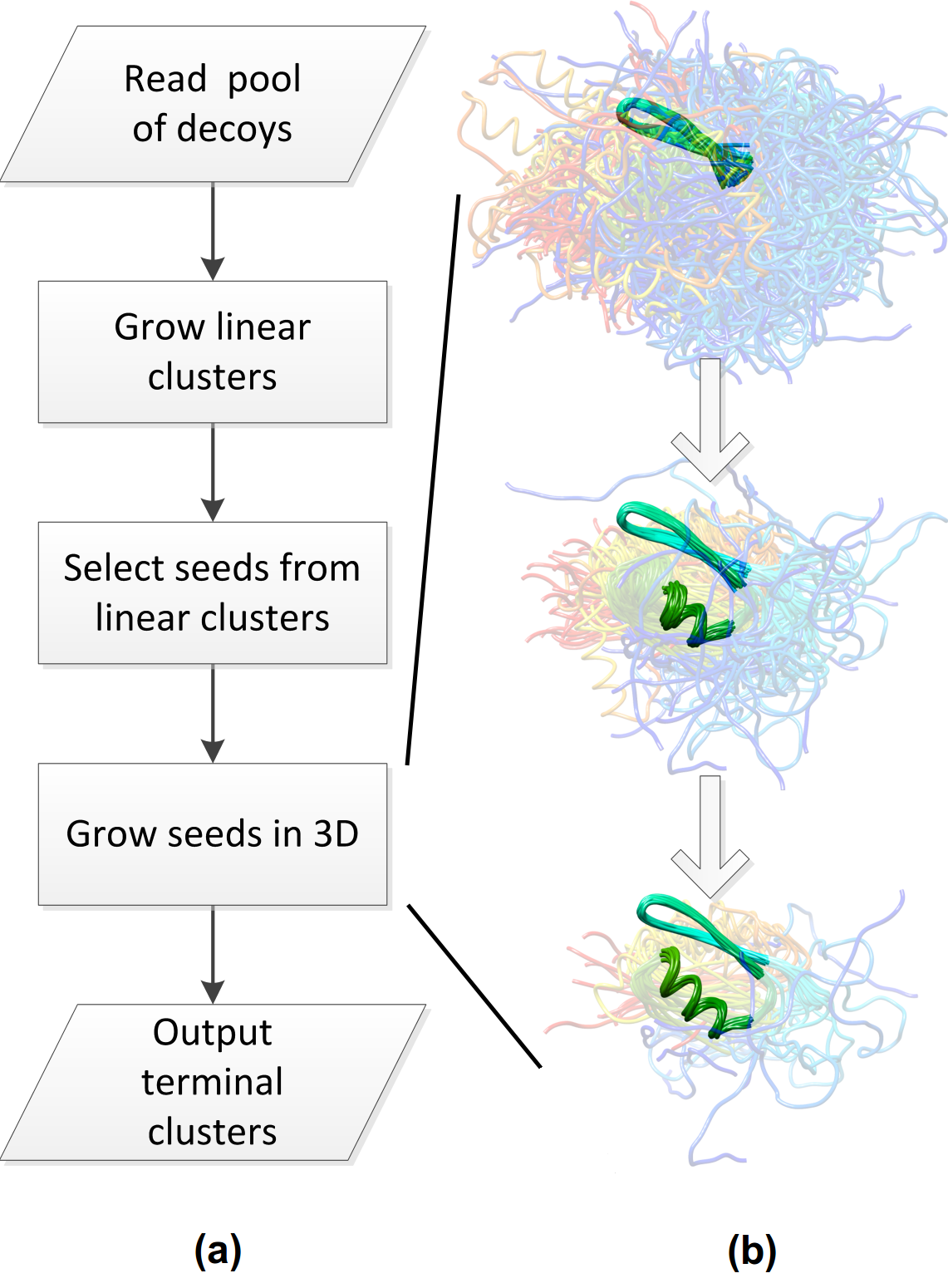}
\caption{(\textbf{a}) Flowchart of the heuristic GC implementation. (\textbf{b}) An example of 3D cluster growing step. The supporting decoys for each step are shown as transparent ribbon diagrams. All structures are coloured by a gradient from blue (N-terminus) to red (C-terminus). Starting from a 16-residue seed the cluster is grown by 8 residues, which are not linear in sequence, then extended by 4 more residues, after which no more suitable candidates found and the procedure is terminated. The illustration is based on the 500 Rosetta decoys produced for a 111 residues long protein target (PDB entry 2C60).}
\end{figure}

\subsection{Overview}
The naive exhaustive search of all terminal clusters would involve costly all-\textit{vs}-all RMSD minimisations, the number of which grows quadratically with the number of models. Moreover, all possible subsets of all residues have to be considered, the number of which grows exponentially with the sequence length. To tackle the computational complexity we split the problem into a two-step procedure.

A short stretch of consecutive residues with similar backbone torsion angles is likely to have a low structural variability in the pool of decoys. We can exploit this fact to quickly grow granules linearly in protein sequence by iteratively merging the initial one-residue clusters (\ref{eq:initial_single}). The results of this step serve as input for the full-scale 3D granular clustering procedure. Hence it can be viewed as preprocessing to reduce the search space. The overall procedure is illustrated in Fig. \ref{fig:01}a.

The implementation of the production rule (\ref{eq:prod}) is split into two parts. Initially, the two input clusters are combined into one, using a simplified production rule:

\begin{equation}
\operatorname{Prod}'(C_1,C_2) = (R_1 \cup R_2, M_1 \cap M_2)
\label{eq:prodprime}
\end{equation}

Afterwards, the problem of alternative conformations and output cluster validity is solved by  \textit{subclustering} the result (\ref{eq:prodprime}). Specifically, let $D(R,M)$ be a matrix $|M| \times |M|$ of distances between the models in $M$ evaluated on residues in $R$. Let $\mathcal{C}(D|h)$ be a standard clustering algorithm parametrised by some constant $h$, such as the number of clusters or a density threshold. The algorithm takes distance matrix $D$ as an input and produces clusters $\{M_1,..M_K\}$. We define subclustering procedure $\operatorname{Sub}(C|D)$ (parameters $v,s,h$ omitted for brevity) of a cluster  $C = (R,M)$ using distance information $D(R,M)$ as follows:

\begin{equation}
\begin{gathered}
\operatorname{Sub}(C|D,v,s,h) = \bigcup_{M_i\in \mathcal{C}(D|h)}\{(R,M_i): \operatorname{Valid}(R,M_i|v,s)\} \\
\bigcup_{i} M_i \subseteq M; M_i \cap M_j = \varnothing, i < j \leq K 
\end{gathered}
\label{eq:subclustering}
\end{equation}

In this implementation, the cluster validity (\ref{eq:valid}) is ensured, while the requirement for cluster saturation (\ref{eq:saturated}) is relaxed and depends on the properties and parameters of the standard clustering algorithm chosen. Note that the definition (\ref{eq:subclustering}) implies deterministic rather than probabilistic cluster assignments, but can be easily generalised for the latter.

\subsection{Linear cluster growing}

Let $\mathbf{S}^0=\bigcup_{1 < k < N}\Big\{(\{k\},\mathbf{M})\Big\}$ be the initial state of one-residue clusters. Each cluster has an associated distance matrix in the space of backbone torsion angles $D_{\phi,\psi}(\{k\},\mathbf{M})$. The first and the last residues are omitted, since they will lack either $\phi$ or $\psi$ by definition.

The successive states $\mathbf{S}^{i+1}$ are constructed by application of the production rule (\ref{eq:prodprime}) with subsequent subclustering (\ref{eq:subclustering}) for every two clusters from $\mathbf{S}^{i}$ that describe adjacent segments by the residue number. Let $C_1 = (R_1,M_1) \in \mathbf{S}^{i}$, $C_2 = (R_2,M_2) \in \mathbf{S}^{i}$. Then their product is
\begin{equation}
\mathbf{P}^{i+1} = \bigcup_{C_1, C_2 \in \mathbf{S}^{i}} \{\operatorname{Prod'}(C_1,C_2): \max{R_1} = \min{R_2}-1 \}
\label{eq:initial_clusters}
\end{equation}
And the next state is given by
\begin{equation}
\mathbf{S}^{i+1} = \bigcup_{C \in \mathbf{P}^{i+1}} \operatorname{Sub}(C|D_{\phi,\psi})
\label{eq:initial_clusters2}
\end{equation}

By definition (\ref{eq:initial_clusters2}), $\mathbf{S}^{i}$ contains clusters of length $2^i$, The procedure will eventually terminate on its own, when either no valid clusters can be produced or $2^i$ surpasses the sequence length. However, structural variability between the supporting models will become less predictable with the cluster length, even if the backbone torsion angles are similar, so imposing a reasonable limit of $i_{max}$ is necessary. Thus the output of the first level of granular clustering is a set of terminal clusters $\mathbf{S}^{i_{max}}$ and a set of intermediate clusters $\mathbf{L} = \bigcup_{0 < i < i_{max}}\{\mathbf{S}^{i}\}$.

\subsection{Seed selection}

Next, we need to select the "seeds" (\textit{i.e.} the initial state) of 3D cluster growing. It would be reasonable to start from the longest segments available, $\mathbf{S}^{i_{max}}$. However, they may overlap by residue numbers and supporting models to a large extent. We have observed that such overlapping clusters, if grown, frequently produce similar end-results (data not shown), and eliminating them would only slightly diminish the total coverage, while greatly reducing the search space. To this end, a non-redundant set of seeds from the longest linear clusters $\mathbf{S}_{seeds} \subseteq \mathbf{S}^{i_{max}}$ is constructed as follows.

Here we use the Jaccard index to quantify similarity of two sets $\operatorname{Jac}(A,B)=\frac{|A \cap B|}{|A \cup B|}$. Two clusters $C_1=(R_1,M_1)$, $C_2=(R_2,M_2)$ are considered independent with respect to parameter $J_{max}$ if the sets of their residues and supporting models have pairwise similarity of at most $J_{max}$:
\begin{equation}
\operatorname{Ind}(C_1, C_2|J_{max}) \iff
\begin{array}{l}
\operatorname{Jac}(R_1,R_2) \leq J_{max}\\
\operatorname{Jac}(M_1,M_2) \leq J_{max}
\end{array}
\end{equation}

A set of clusters $\mathbf{S}=\{(R_1,M_1),...,(R_k,M_k)\}$ is considered independent with respect to $J_{max}$ if all possible pairs of clusters in the set are independent with respect to $J_{max}$:

\begin{equation}
\operatorname{Ind}(\mathbf{S}|J_{max}) \iff
\operatornamewithlimits{Ind}_{\forall C_i, C_j \in \mathbf{S}, i \neq j}(C_i,C_j|J_{max})
\end{equation}

Let $\mathbf{S}^{i_{max}}_{sorted}=\Big\langle C_1=(R_1,M_1),...,C_k=(R_k,M_k)\Big\rangle$ be a sequence of clusters from  $\mathbf{S}^{i_{max}}$ ordered by decreasing support, i.e. $|M_{j-1}|\geq|M_{j}|, j=2..k$. A set of seeds $\mathbf{S}_{seeds}$ is defined as follows:
\begin{enumerate}
\item The cluster with the largest support is included into the seeds set: $C_1 \in \mathbf{S}_{seeds}$.
\item Each subsequent cluster from $\mathbf{S}^{i_{max}}_{sorted}$ is included into the seeds set if it does not break the independence criterion with the already included clusters: 
\begin{equation}
C_i \in \mathbf{S}_{seeds} \iff \operatorname{Ind}\Big( \bigcup_{j < i}\{C_j: C_j \in \mathbf{S}_{seeds}\} \cup C_i \Big)
\end{equation}
\end{enumerate}

\subsection{3D cluster growing} 
At the start we have the set of seeds $\mathbf{S}_{seeds}$ and the set of linear clusters $\mathbf{L}$, grouped by their lengths 2 to $2^{i_{max}-1}$. Here we search for clusters that are similar in 3D (but may be non-linear in sequence), using the distance matrix $D_{rmsd}(R,M)$ defined by pairwise RMSD of atomic coordinates between all models in $M$ on residues in $R$.

\begin{algorithm}[tb]
\begin{algorithmic}[1]
    \Require $C_{seed}\in \mathbf{S}_{seeds}, \mathbf{L}=\{\mathbf{S}^1,...,\mathbf{S}^{i_{max}-1}\};$
    \Procedure{GrowCluster}{$C_{seed},\mathbf{L}$}
    \State $C_{cur} := C_{seed}$
    \State $i_{cur} \gets i_{max}-1$
    \Repeat
        \State $C_{cand} \gets \operatorname{Select}(C_{cur},\mathbf{S}^{i_{cur}})$
        \Comment{Select the best extension}
		\If{$C_{cand} = \varnothing$}
			\State $i_{cur} \gets i_{cur}-1$
		\Else
			\State $C_{cur} \gets C_{cand}$
        \EndIf
    \Until{$i_{cur}=0$}
    \Comment{No more candidate clusters}
    \State \Return{$C_{cur}$}
    \EndProcedure
\end{algorithmic}
\caption{Cluster growing}
\label{alg:growing}
\end{algorithm}

The high-level algorithm for growing a cluster from a seed is presented in Algorithm \ref{alg:growing}. The procedure $\operatorname{Select}(C,\mathbf{S})$ is a greedy heuristic for cluster extension and involves the following computations. First, we produce all possible candidate extensions of a seed $C \in \mathbf{S}_{seeds}$ with the given set of clusters $\mathbf{S} \in \mathbf{L}$ using production rule (\ref{eq:prodprime}):
\begin{equation}
\operatorname{Combine}(C,\mathbf{S}) = \bigcup_{S\in \mathbf{S}} \operatorname{Prod}'(C,S)
\end{equation}
Then the candidate extensions are subclustered (\ref{eq:subclustering}) and the results are aggregated into one set:
\begin{equation}
\operatorname{Split}(C,\mathbf{S}) = \bigcup_{C'\in \operatorname{Combine}(C,\mathbf{S})} \operatorname{Sub}(C'|D_{rmsd}(C'))
\end{equation}
Finally, the best-scoring subcluster is selected:
\begin{equation}
\label{eq:select}
\operatorname{Select}(C,\mathbf{S}) = 
\argmin_{C'\in \operatorname{Split}(C,\mathbf{S})} V(C'),
\end{equation}
where $V(C)$ is a cluster scoring function. In case the folding algorithm provides scores for individual decoys, such as Rosetta energy function (\citealp{rohl2004protein}), they may be included in the evaluation of the clusters. The scoring functions are generally designed to provide prediction of the likeliness to the native structure and could therefore be helpful towards estimating the quality of a resulting cluster. We define $V(C)$ as sum of Rosetta energy scores $E$ ($E=-1$ if energies are not available) of the decoys in a cluster $C$ divided by the average pairwise superposition RMSD $\mathcal{R}(C)$ of the residues included in the cluster.
\begin{equation}
\label{eq:variability}
V(C)=\frac{\sum_{i \in M}E_i}{\mathcal{R}(C)+1}
\end{equation}

\section{Implementation}
The GC algorithm is implemented as a Python script. Biopython (\citealp{Cock:Biopython}) is used to process the Protein Data Bank (PDB) files. Structure superpositions and RMSD calculations are done with PyRMSD (\citealp{Gil:PyRMSD}). Subclustering is performed with the mean-shift (MS) algorithm (\citealp{comaniciu2002mean}). Backbone torsion angles are used directly as samples for MS, while all-\textit{vs}-all RMSD matrices are firstly embedded into a 2D space with multidimensional scaling (MDS). Standard MS and MDS implementations from Scikit-learn (\citealp{scikit-learn}) package are used. 

The algorithm uses a number of parameters that affect the end-result in various ways (Table \ref{tab:params}). The default values indicated were found to be a reasonable starting point for problem-specific fine-tuning.

\begin{table}
\centering
\caption{GC algorithm parameters.} 
\label{tab:params}
\begin{tabular}{lp{5cm}p{5cm}p{1.8cm}}\toprule 
  \textbf{Parameter} & \textbf{Description} & \textbf{Affects} & \textbf{Default} \\\midrule
  $h_{lin}$ & Mean-shift bandwidth for subclustering in the space of torsion angles &  Precision of linear clusters & 1.2 \\
  \\
  $s_{lin}$ & Minimal support for the linear cluster growing & Number of linear clusters & 10\% of the pool\\
  \\
  $i_{max}$ & Number of repeats of the cluster doubling procedure & Maximal length of seeds & 4\\
  \\
  $J_{max}$ & Maximal overlap between seed candidates& Number of seeds & 0.5\\
  \\
  $h_{rmsd}$ & Mean-shift bandwidth for subclustering in the space of RMSD distances & Precision of output clusters & 0.5 \\
  \\
  $s_{rmsd}$ & Minimal support for the 3D cluster growing & Length of output clusters & 10 \\\bottomrule
\end{tabular}
\end{table}

\section{Results}

A test dataset of 295 crystal structures (\citealp{Bibby:tz5014}) was used to evaluate the performance of the GC algorithm. 500 decoys for each target were folded using Rosetta (\citealp{leaver2011rosetta3}). We compared the partial protein clusters produced by the GC algorithm using default parameters to the clusters generated by the \texttt{AMPLE} pipeline (\citealp{Bibby:tz5014}). This pipeline contains a cluster-and-truncate component which is the most comparable method result-wise, while conceptually being the opposite, since the clusters are generated by global alignment and elimination of diverging segments, with subsequent re-clustering. Additionally, we have compared the relative success of the MR search models provided by the two methods.

\subsection{Coverage}

To evaluate the clusters quality \textit{per se}, we estimated their \textit{coverage} (in terms of fraction of the total sequence length) as a function of the maximal allowed RMSD from the true structure. Let average RMSD of models in the cluster $(R,M)$ with respect to the true experimental structure $\mathbf{N}$ be given by $\mathcal{R}(R,M|\mathbf{N})$. For a set of output clusters $\mathbf{C}^{O}=\{(R_1,M_1),...,(R_{N_{O}},M_{N_{O}})\}$ a subset of clusters that are within an RMSD of $r$ from the true structure is given by
\begin{equation}
\widetilde{\mathbf{C}}^{O}(\mathbf{N}|r) = \{(R_i,M_i)\in \mathbf{C}^{O}, \mathcal{R}(R_i,M_i|\mathbf{N})\leq r\}
\end{equation}
We will consider the coverage of a single 'best' cluster (understood as the cluster containing the largest number of residues) from this subset
\begin{equation}
\operatorname{Cov}_{s}(\mathbf{C}^{O},\mathbf{N}|r)=\ddfrac{\max_{i}\{| R_i \cap \mathbf{N}|: (R_i,M_i) \in \widetilde{\mathbf{C}}^{O}(\mathbf{N}|r)\}}{|\mathbf{N}|}
\label{eq:coverage_max}
\end{equation}
We also define the total coverage of this subset as
\begin{equation}
\operatorname{Cov}_{t}(\mathbf{C}^{O},\mathbf{N}|r)=\ddfrac{\Big|\bigcup_{i} \{R_i \cap \mathbf{N}: (R_i,M_i) \in \widetilde{\mathbf{C}}^{O}(\mathbf{N}|r)\}\Big|}{|\mathbf{N}|}
\label{eq:coverage}
\end{equation}

\begin{figure}
\centering
\label{fig:02}
\includegraphics[width=0.5\textwidth]{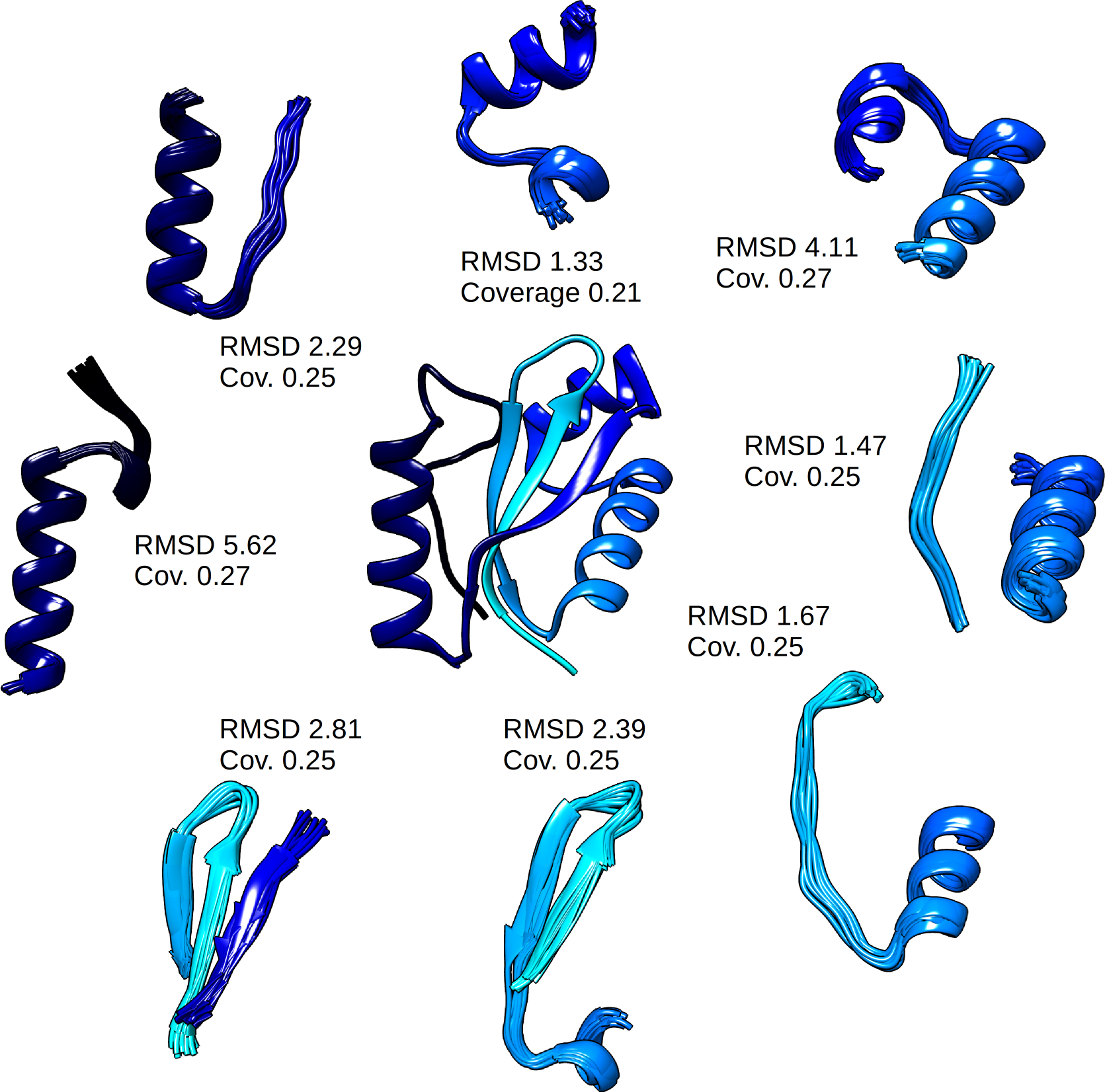}
\caption{Ribbon diagrams of a 97 residues long protein target in the centre (PDB entry 1MK0) and sample clusters obtained by GC containing 10 models each. All structures are coloured by a gradient from light blue (N-terminus) to dark blue (C-terminus). For each cluster, the average C$\alpha$ RMSD relative to the crystal structure and coverage of the cluster are indicated.}
\end{figure}

A set of clusters output by the GC algorithm for a protein target, their RMSD to the true structure and coverage are presented in Fig. \ref{fig:02}. As can be seen, all regions are covered with variable precision, sometimes offering different conformations for the same segment. It is worth mentioning that \texttt{AMPLE} clusters (Supplementary Fig. S1, S2) follow a completely different pattern: starting from a small core of structurally conserved residues they gradually increase in coverage while becoming more divergent.

\begin{figure}
\centering
\includegraphics[width=1\textwidth]{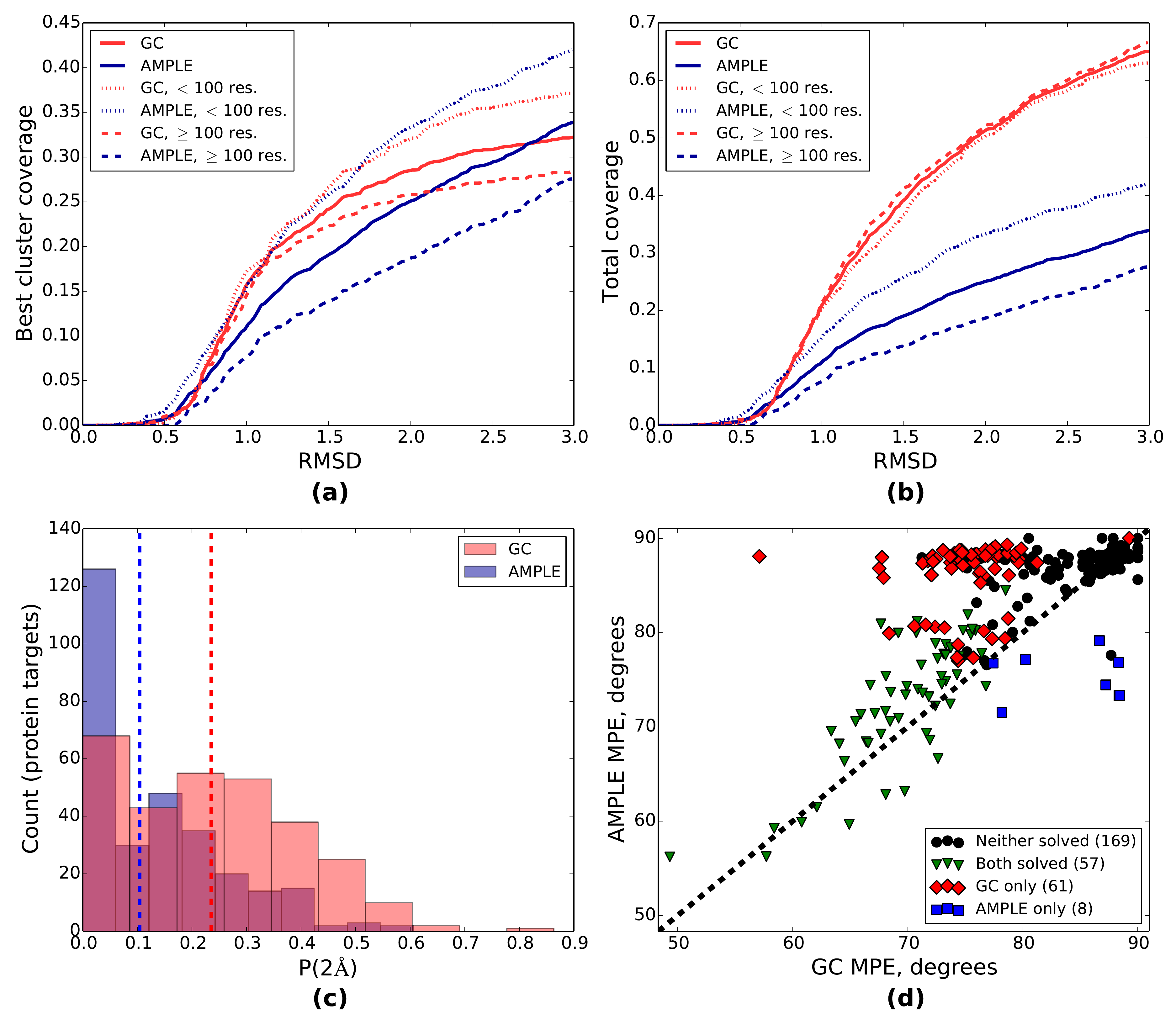}
\caption{Evaluation of the GC algorithm. (\textbf{a}) Best cluster coverage as a function of the RMSD value to the true structure (\ref{eq:coverage_max}). The values averaged over the benchmark set of 295 target structures are plotted. \texttt{AMPLE} results plotted in blue, GC -- in red. Solid line gives mean values for the entire dataset, dotted line represents proteins with sequence length less than a 100 residues, dashed line -- proteins 100 residues and longer. (\textbf{b}) The corresponding total coverage function (\ref{eq:coverage}). (\textbf{c}) Histogram of integral coverage (\ref{eq:integral_coverage}), calculated for all clusters produced by \texttt{AMPLE} (blue) and GC (red) that are within 2\r{A} RMSD to the true structure. The median values are shown by vertical dashed lines: 0.1 for \texttt{AMPLE}, 0.24 for GC. (\textbf{d}) Scatter plot comparing the results of the MR search using clusters obtained through both algorithms. Each point corresponds to a target from the benchmark set. The x and y axes give the minimal MPE values among the MR solutions for this target obtained with GC clusters and \texttt{AMPLE} clusters respectively. Points above the diagonal represent structures where GC clusters yielded MR solutions closer to the true structure than the \texttt{AMPLE} clusters, while the opposite is true for the points below the diagonal. Colour-coding indicates the outcome of automatic model rebuilding in \texttt{SHELXE} of the respective MR solutions: failed with both \texttt{AMPLE} and GC clusters (black), succeeded with both \texttt{AMPLE} and GC clusters (green), succeeded with GC, but not with \texttt{AMPLE} (red), succeeded with \texttt{AMPLE}, but not GC (blue).}
\label{fig:03}
\end{figure}

Averaging of the $\operatorname{Cov}_{s}(r)$ and $\operatorname{Cov}_{t}(r)$ functions over the benchmark set of 295 target structures can give an idea about the comparative performance of the two methods (Fig. \ref{fig:03}a,b). In terms of the best coverage by a single cluster, GC procedure clearly outperforms the \texttt{AMPLE} routine for all reasonably precise clusters (up to 2.5\r{A} RMSD from the target structure), with the advantage even more pronounced for longer sequences. In addition, the collective coverage by all clusters output by the GC algorithm is also much better, exceeding that of \texttt{AMPLE} by a factor of two for almost any RMSD cut-off (Fig. \ref{fig:03}b). The sequence length has a negligible effect on the total coverage (dotted and dashed lines on Fig. \ref{fig:03}b). This implies that longer protein chain modelled by Rosetta can still produce useful results if local clusters are our target.

In addition, to assess the distribution of cluster quality for individual targets, we have analysed the integral coverage function $P(\mathbf{C}^{O},\mathbf{N}|r_{max})$, which is analogous to the area under curve for a receiver operating characteristic (\citealp{fawcett2006introduction}). The maximal acceptable level of RMSD deviation $r_{max}$ is used to bring the metric to the scale of 0 to 1. Let $\langle r_{1},...,r_{N_O}\rangle$ be a sequence of cluster RMSDs to the native structure $\mathcal{R}(R_i,M_i|\mathbf{N})$ sorted from smallest to largest, i.e. $r_{i}\leq r_{i+1}$. This integral coverage function is calculated using a trapezoidal formula:
\begin{multline}
P(\mathbf{C}^{O},\mathbf{N}|r_{max}) = \frac{1}{2r_{max}}  \sum_{k=1}^{N-1} \left( r_{k+1} - r_{k} \right) \times\\ 
\times \left( \operatorname{Cov}_{t}(\mathbf{C}^{O},\mathbf{N}|r_{k+1}) +  \operatorname{Cov}_{t}(\mathbf{C}^{O},\mathbf{N}|r_{k}) \right),
\label{eq:integral_coverage}
\end{multline}
where $N$ is the number of clusters with $r_{k} \leq r_{max}$, and $\operatorname{Cov}_{t}(r_{k})$ -- total coverage at $r_{k}$ as defined by (\ref{eq:coverage}).

Fig. \ref{fig:03}c shows comparison of $P$ distributions calculated on the test set for $r_{max}=2\textrm{\r{A}}$ (see also Supplementary Fig. S3). Here again the superior capabilities of the GC algorithm are very apparent, with more than two-fold increase in median integral coverage compared with the \texttt{AMPLE} output.
Overall the statistics presented indicate the strength of the local bottom-up approach used in GC. 

\subsection{Use for molecular replacement}
We have also evaluated the usefulness of GC-based partial models towards phasing crystal structures of proteins by MR on the same test set of 295 crystal structures of non-homologous proteins (\citealp{Bibby:tz5014}). Previously these authors have explored the use of \texttt{AMPLE}-derived clusters towards phasing this test set by first running the MR procedure and thereafter attempting structure rebuilding and extension of this solution using \texttt{SHELXE} (\citealp{Thorn:shelxemr}), repeating the whole calculation for every cluster. Here we have employed a further modification of this routine, which allows testing of the clusters' performance in a radically reduced computational time. Initially, for each target structure an MR search using the obtained clusters as search models was performed by \texttt{Phaser} (\citealp{McCoy:phaser}). At this point, all MR solutions obtained with various clusters for a given target were evaluated with respect to the similarity to the true crystal structure. To this end, calculation of their mean phase error (MPE) with respect to the true structure was performed using \texttt{cphasematch} (\citealp{winn2011overview}). Thereafter, only the MR solution yielding the lowest MPE underwent structure rebuilding. A case was considered solved if this procedure could advance beyond a certain minimal chain length and correlation between the rebuilt model and electron density, as evaluated in \texttt{SHELXE}. Further details as well as a table with complete results are provided in the Supplement. Fig. \ref{fig:03}d summarises the performance of GC and \texttt{AMPLE} clusters as MR search models. In more than two thirds of the test cases lower MPE values could be achieved with the GC clusters. The value of 80\textdegree{ }appears to be a cut-off beyond which the automatic model extension is unlikely to succeed, typically because the MR solution has been completely wrong in the first place. In 144 cases, the minimal MPE of the MR solutions obtained with GC clusters was below the said cut-off, compared to 70 such cases with the \texttt{AMPLE} clusters. A majority of these solutions could be successfully rebuilt and expanded in \texttt{SHELXE}. Ultimately, only 65 of the 295 test cases could be successfully phased using \texttt{AMPLE} clusters, while 118 structures could be phased with GC models. The use of GC-based models has thus resulted in an about two-fold higher success rate of the MR procedure.

Of further note, the granular approach to search model generation frequently results in clusters that do not overlap by sequence. This means that during the MR procedure one can attempt to place two or more independent search models at once. This approach has allowed us to obtain a correct MR solution in at least one additional case (1SBX, data not shown). 
\section{Discussion}

Here we have proposed a novel method to produce clusters of partial models from protein decoys based on local structural similarity, which falls under the granular computing paradigm. It should be noted that the existing methods of protein model clustering typically consider a full-length decoy as a single data point; all such decoys are then clustered upon some sort of a single-alignment procedure. This imposes an obvious limitation on the clustering algorithm, since one can not operate with less than a whole decoy.  In contrast, the GC approach operates in a much larger search space, since the decoy data are initially granulated down to the level of a single residue. As we have shown here, this enables the design of a clustering algorithm that is very efficient in extracting the structural information from a pool of \textit{de-novo} modelled decoys. While more demanding computationally compared to approaches based on a single structural alignment, GC is nevertheless capable of yielding extremely useful results for typical proteins even when using modest computational resources.

Our implementation is the first 'proof-of-concept' of the GC approach to protein structures, and application of more advanced heuristic search strategies is likely to follow. Moreover, we envisage further development of this approach towards a range of research questions. In particular, this could include algorithms to detect non-linear structural motifs in a large set of 3D structures, such as the experimental structures available in the PDB. In addition, by incorporating amino acid sequence distribution in the observed clusters, one could obtain variable-length fragment library for protein structure prediction. In this case, fragments with long-distance interactions could be used for the generation of prior spatial constraints to be utilized during the \textit{ab initio} protein modelling.

In conclusion, we have developed an alternative view of the structural protein clustering problem, which enables 'growing' clusters of partial models from local similarities observed in sampled conformations. We have shown that solving the phase problem in X-ray crystallography is an area that can immediately benefit from the results obtained here. We hope they will facilitate development of further novel techniques in protein structure prediction as well as aid in experimental structure determination. 

\bibliographystyle{natbib}
\bibliography{localsim}

\begin{thebibliography}{}

\bibitem[Bibby {\em et~al.}(2012)Bibby, Keegan, Mayans, Winn, and
  Rigden]{Bibby:tz5014}
Bibby, J., Keegan, R.~M., Mayans, O., Winn, M.~D., and Rigden, D.~J. (2012).
\newblock {{AMPLE}: a cluster-and-truncate approach to solve the crystal
  structures of small proteins using rapidly computed ab initio models}.
\newblock {\em Acta Crystallographica Section D\/}, {\bf 68}(12), 1622--1631.

\bibitem[Cock {\em et~al.}(2009)Cock, Antao, Chang, Chapman, Cox, Dalke,
  Friedberg, Hamelryck, Kauff, Wilczynski, and de~Hoon]{Cock:Biopython}
Cock, P. J.~A., Antao, T., Chang, J.~T., Chapman, B.~A., Cox, C.~J., Dalke, A.,
  Friedberg, I., Hamelryck, T., Kauff, F., Wilczynski, B., and de~Hoon, M.
  J.~L. (2009).
\newblock Biopython: freely available {P}ython tools for computational
  molecular biology and bioinformatics.
\newblock {\em Bioinformatics\/}, {\bf 25}(11), 1422--1423.

\bibitem[Comaniciu and Meer(2002)Comaniciu and Meer]{comaniciu2002mean}
Comaniciu, D. and Meer, P. (2002).
\newblock Mean shift: A robust approach toward feature space analysis.
\newblock {\em IEEE Transactions on pattern analysis and machine
  intelligence\/}, {\bf 24}(5), 603--619.

\bibitem[Das and Baker(2008)Das and Baker]{das2008macromolecular}
Das, R. and Baker, D. (2008).
\newblock Macromolecular modeling with {R}osetta.
\newblock {\em Annu. Rev. Biochem.}, {\bf 77}, 363--382.

\bibitem[Fawcett(2006)Fawcett]{fawcett2006introduction}
Fawcett, T. (2006).
\newblock An introduction to {ROC} analysis.
\newblock {\em Pattern recognition letters\/}, {\bf 27}(8), 861--874.

\bibitem[Gil and Guallar(2013)Gil and Guallar]{Gil:PyRMSD}
Gil, V.~A. and Guallar, V. (2013).
\newblock py{RMSD}: a {P}ython package for efficient pairwise {RMSD} matrix
  calculation and handling.
\newblock {\em Bioinformatics\/}, {\bf 29}(18), 2363--2364.

\bibitem[Kabsch(1976)Kabsch]{kabsch1976solution}
Kabsch, W. (1976).
\newblock A solution for the best rotation to relate two sets of vectors.
\newblock {\em Acta Crystallographica Section A: Crystal Physics, Diffraction,
  Theoretical and General Crystallography\/}, {\bf 32}(5), 922--923.

\bibitem[Leaver-Fay {\em et~al.}(2011)Leaver-Fay, Tyka, Lewis, Lange, Thompson,
  Jacak, Kaufman, Renfrew, Smith, Sheffler, {\em et~al.}]{leaver2011rosetta3}
Leaver-Fay, A., Tyka, M., Lewis, S.~M., Lange, O.~F., Thompson, J., Jacak, R.,
  Kaufman, K., Renfrew, P.~D., Smith, C.~A., Sheffler, W., {\em et~al.} (2011).
\newblock {ROSETTA3}: an object-oriented software suite for the simulation and
  design of macromolecules.
\newblock {\em Methods in enzymology\/}, {\bf 487}, 545.

\bibitem[Malashkevich {\em et~al.}(2015)Malashkevich, Higgins, Almo, and
  Lai]{malashkevich2015switch}
Malashkevich, V.~N., Higgins, C.~D., Almo, S.~C., and Lai, J.~R. (2015).
\newblock A switch from parallel to antiparallel strand orientation in a
  coiled-coil {X}-ray structure via two core hydrophobic mutations.
\newblock {\em Peptide Science\/}, {\bf 104}(3), 178--185.

\bibitem[Mariani {\em et~al.}(2013)Mariani, Biasini, Barbato, and
  Schwede]{Mariani01112013}
Mariani, V., Biasini, M., Barbato, A., and Schwede, T. (2013).
\newblock {lDDT}: a local superposition-free score for comparing protein
  structures and models using distance difference tests.
\newblock {\em Bioinformatics\/}, {\bf 29}(21), 2722--2728.

\bibitem[McCoy {\em et~al.}(2007)McCoy, Grosse-Kunstleve, Adams, Winn, Storoni,
  and Read]{McCoy:phaser}
McCoy, A.~J., Grosse-Kunstleve, R.~W., Adams, P.~D., Winn, M.~D., Storoni,
  L.~C., and Read, R.~J. (2007).
\newblock {Phaser crystallographic software}.
\newblock {\em Journal of Applied Crystallography\/}, {\bf 40}(4), 658--674.

\bibitem[Pedregosa {\em et~al.}(2011)Pedregosa, Varoquaux, Gramfort, Michel,
  Thirion, Grisel, Blondel, Prettenhofer, Weiss, Dubourg, Vanderplas, Passos,
  Cournapeau, Brucher, Perrot, and Duchesnay]{scikit-learn}
Pedregosa, F., Varoquaux, G., Gramfort, A., Michel, V., Thirion, B., Grisel,
  O., Blondel, M., Prettenhofer, P., Weiss, R., Dubourg, V., Vanderplas, J.,
  Passos, A., Cournapeau, D., Brucher, M., Perrot, M., and Duchesnay, E.
  (2011).
\newblock Scikit-learn: Machine learning in {P}ython.
\newblock {\em Journal of Machine Learning Research\/}, {\bf 12}, 2825--2830.

\bibitem[Pedrycz and Bargiela(2002)Pedrycz and Bargiela]{Pedrycz:granular}
Pedrycz, W. and Bargiela, A. (2002).
\newblock Granular clustering: a granular signature of data.
\newblock {\em Systems, Man, and Cybernetics, Part B: Cybernetics, IEEE
  Transactions on\/}, {\bf 32}(2), 212--224.

\bibitem[Rohl {\em et~al.}(2004)Rohl, Strauss, Misura, and
  Baker]{rohl2004protein}
Rohl, C.~A., Strauss, C.~E., Misura, K.~M., and Baker, D. (2004).
\newblock Protein structure prediction using {R}osetta.
\newblock {\em Methods in enzymology\/}, {\bf 383}, 66--93.

\bibitem[Rossmann(1990)Rossmann]{rossmann1990molecular}
Rossmann, M.~G. (1990).
\newblock The molecular replacement method.
\newblock {\em Acta Crystallographica Section A: Foundations of
  Crystallography\/}, {\bf 46}(2), 73--82.

\bibitem[Shortle {\em et~al.}(1998)Shortle, Simons, and Baker]{Shortle15091998}
Shortle, D., Simons, K.~T., and Baker, D. (1998).
\newblock Clustering of low-energy conformations near the native structures of
  small proteins.
\newblock {\em Proceedings of the National Academy of Sciences\/}, {\bf
  95}(19), 11158--11162.

\bibitem[Thorn and Sheldrick(2013)Thorn and Sheldrick]{Thorn:shelxemr}
Thorn, A. and Sheldrick, G.~M. (2013).
\newblock {Extending molecular-replacement solutions with {SHELXE}}.
\newblock {\em Acta Crystallographica Section D\/}, {\bf 69}(11), 2251--2256.

\bibitem[Winn {\em et~al.}(2011)Winn, Ballard, Cowtan, Dodson, Emsley, Evans,
  Keegan, Krissinel, Leslie, McCoy, {\em et~al.}]{winn2011overview}
Winn, M.~D., Ballard, C.~C., Cowtan, K.~D., Dodson, E.~J., Emsley, P., Evans,
  P.~R., Keegan, R.~M., Krissinel, E.~B., Leslie, A.~G., McCoy, A., {\em
  et~al.} (2011).
\newblock Overview of the {CCP4} suite and current developments.
\newblock {\em Acta Crystallographica Section D: Biological Crystallography\/},
  {\bf 67}(4), 235--242.

\bibitem[Yao {\em et~al.}(2013)Yao, Vasilakos, and
  Pedrycz]{Jing:granularreview}
Yao, J.~T., Vasilakos, A., and Pedrycz, W. (2013).
\newblock Granular computing: Perspectives and challenges.
\newblock {\em Cybernetics, IEEE Transactions on\/}, {\bf 43}(6), 1977--1989.

\bibitem[Zemla(2003)Zemla]{Zemla01072003}
Zemla, A. (2003).
\newblock {LGA}: a method for finding {3D} similarities in protein structures.
\newblock {\em Nucleic Acids Research\/}, {\bf 31}(13), 3370--3374.

\bibitem[Zhang(2009)Zhang]{zhang2009protein}
Zhang, Y. (2009).
\newblock Protein structure prediction: when is it useful?
\newblock {\em Current opinion in structural biology\/}, {\bf 19}(2), 145--155.

\bibitem[Zhang(2014)Zhang]{Zhang:Tasser}
Zhang, Y. (2014).
\newblock Interplay of {I-TASSER} and {QUARK} for template-based and ab initio
  protein structure prediction in {CASP10}.
\newblock {\em Proteins: Structure, Function, and Bioinformatics\/}, {\bf 82},
  175--187.

\bibitem[Zhang and Skolnick(2004)Zhang and Skolnick]{zhang2004spicker}
Zhang, Y. and Skolnick, J. (2004).
\newblock {SPICKER}: A clustering approach to identify near-native protein
  folds.
\newblock {\em Journal of computational chemistry\/}, {\bf 25}(6), 865--871.

\end{thebibliography}
\end{document}